\documentclass[showpacs,twocolumn,aps,superscriptaddress]{revtex4}
\usepackage{color}

\usepackage{amsmath}
\usepackage{bm}
\usepackage{epsfig}
\usepackage{amssymb}

\usepackage{graphicx}
\usepackage{epsfig}
\usepackage{psfrag}
\newcommand{\dd}{\text{d}}

\newcommand{\ee}{\text{e}}

\newcommand{\p}{\partial}
\newcommand{\ppt}{\frac{\partial}{\partial t}}

\newcommand{\D}{\mathcal{D}}

\newcommand{\hs}{h_\mathrm{s}}
\newcommand{\ms}{m_\mathrm{s}}

\begin{document}

\title{Second-order dynamic transition in a $p=2$ spin-glass model} 

\author{Kristina van Duijvendijk}
\affiliation{Laboratoire Mati\`ere et Syst\`emes Complexes (CNRS UMR
  7057), Universit\'e Paris Diderot--Paris 7, 10 rue Alice Domon et L\'eonie
  Duquet, 75205 Paris Cedex 13, France}

\author{Robert L. Jack}
\affiliation{Department  of Physics, University of Bath, Bath, BA2 7AY, 
United Kingdom}

\author{ Fr\'ed\'eric van Wijland}
\affiliation{Laboratoire Mati\`ere et Syst\`emes Complexes (CNRS UMR
  7057), Universit\'e Paris Diderot--Paris 7, 10 rue Alice Domon et L\'eonie
  Duquet, 75205 Paris Cedex 13, France}

\begin{abstract}
We consider the dynamics of a disordered $p$-spin model with $p=2$,
analyzing the dynamics within Ruelle's thermodynamic formalism, 
We use an indicator of the dynamical activity to construct the relevant 
dynamical Gibbs ensemble.  We prove that the dynamics in the low-temperature
(spin glass) phase of the model take place at a second-order 
phase transition between dynamically active and inactive trajectories.
We also show that the same behaviour is found in a related model
of a three-dimensional ferromagnet.
\end{abstract}

\pacs{75.10.Nr,05.40.-a,64.70.qj}

\maketitle

\section{Introduction}

Glassy systems are characterised by their dynamical properties:
at their glass transitions, they fall out of equilibrium
 on experimental time-scales, and exhibit aging phenomena.  As the glass transition
is approached, their relaxation times increase in a super-Arrhenius  fashion and
the decay of their equilibrium dynamical correlation functions is slower than exponential~\cite{glass}.
In the last fifteen years, several approaches have been developed to 
recover theoretically
the experimental, out-of-equilibrium results (see, for
example, \cite{cugl}).

One of the most striking experimental features of a glass-forming liquid is
that the increase in relaxation time near the glass transition does 
not seem to be accompanied by any significant changes in the liquid 
structure.
However, experiments and computer simulations~\cite{DH-review,
exp-science,exp-colloid,dh-comp} 
both indicate that a {\it dynamical} length scale is growing 
as the glass transition
is approached.  That is, glassy materials are made up of active and inactive
regions of space-time,
namely {\it dynamical heterogeneities}. Based upon these
results, the idea that the glassy properties of a system arise directly from 
their dynamical heterogeneity was developed in \cite{garr,merolle},
stimulating further theoretical understandings of glassy dynamics.  

Spin glasses are magnetic spin systems which exhibit several
features in common with glass-forming liquids.  They are modelled
by spins with quenched random interactions between them,
and have been  extensively investigated both experimentally and 
theoretically (see \cite{rev} for a review). 
The purpose of this paper is to show that the 
glassy dynamics of a particular spin glass model can be understood in terms of
the histories it follows in configuration space.  To this end, we employ
the thermodynamic formalism of histories, developed by Ruelle and coworkers
\cite{ruelle} within the framework of dynamical systems theory, and summarized 
in \cite{thermo} in the context of Markov
dynamics.  While equilibrium statistical mechanics
is concerned with fluctuations in the configuration space of the system, 
Ruelle's formalism focuses on the trajectories (histories) by 
which the system evolves through configuration space.  
The method has been applied recently to kinetically
constrained models of glass-formers~\cite{lettres} and to a Lennard Jones
binary mixture~\cite{ka-science}.
Both these studies distinguish {\it active}
and {\it inactive} histories of the systems, according to the range
of configuration space visited during the history.  In the
kinetically constrained models, it was proven that the active
and inactive histories form distinct populations.
In the language of the thermodynamic formalism, they are separated by a
first-order phase transition in trajectory space.
In Refs.~\cite{lettres,ka-science}, it was argued that 
the heterogeneous dynamics of those models is intrinsically linked
to this transition.  

In this article, we focus on a soft $p$-spin model with $p=2$, 
whose static and dynamic properties can be studied analytically. 
The $p\rightarrow \infty$ limit of the $p$-spin
model, namely the disordered Random Energy Model, was recently 
shown to possess a
connection between the activity of the histories it follows and the 
dynamical heterogeneities in its glassy phase \cite{rem}. 
(Of course, in $p$-spin
models with infinite-ranged interactions, the
dynamical correlation lengths associated with dynamical 
heterogeneity are ill-defined.  
However, the presence of large dynamical
fluctuations in these mean-field models is naturally linked to dynamical
heterogeneity in their finite-dimensional counterparts.  We show that
the $p=2$ spin-glass model is closely related to a three-dimensional 
ferromagnet in which such length scales can be calculated.)
In both the spin glass and ferromagnetic models, we demonstrate the existence 
of a phase transition in trajectory space, offering further evidence 
that these phase transitions are very generally associated with glassy 
systems.

The outline of the paper is as follows: in Section~\ref{sec:model}
we describe the models we will consider and the methods that
we will use.  In section~\ref{sec:results} we construct
a `dynamic phase diagram' that describes the behaviour of the system
in trajectory space.  We interpret our results in 
Sec.~\ref{sec:interp}, discussing the links between the large deviations
that we have derived and the more familiar features of the soft
spin models, and identifying directions for further study.

\section{Model and formalism} 
\label{sec:model}

\subsection{Spin glass model}


We consider a system of $N$ continuous spins $\sigma_i$ whose Hamiltonian 
$\mathcal H$ is given by:
\begin{equation}\label{ham}
\beta \mathcal{H}=-\frac{\beta}{2N}\sum_{i,j}J_{ij}\sigma_i \sigma_j +\frac{u}{N}\sum_{i,j}\sigma_i^2
\sigma_j^2 \;.
\end{equation} 
Here, $\beta=1/T$ as usual, where $T$ is the temperature
and we have set Boltzmann's constant to unity, 
we take $u>0$, 
and the random couplings are Gaussian distributed
\begin{equation}
p(J_{ij})=\frac{1}{\sqrt{2\pi}}\exp\left(-\frac{J_{ij}^2}{2}\right)\;.
\end{equation}

The role of the term proportional to $u>0$ is to suppress configurations with 
extreme values of the spins.  The model is similar to the 
$p$-spin models discussed in~\cite{kac,kirk}.
Like the
spherical $p$-spin model of~\cite{kac} and in contrast to
that of~\cite{kirk}, the model under consideration here can be solved
exactly.  
However, we use the $u$-term instead of a spherical constraint
since it facilitates studies of large-deviations of the activity.  (In 
particular, we note that fluctuations of extensive quantities require
careful treatment in spherical models~\cite{Sollich-sph}, in
which fluctuations of the Lagrange multiplier for the constraint
must be considered.)  
In any case, all of these soft $p$-spin models exhibit finite-temperature `glass
transitions' at which ergodicity is broken~\cite{kirk,crisa,cugliandolodean,
Kost}.
Connections with the structural glass problem have been discussed 
in~\cite{kirk,thir,woly}.
The case of $p=2$ differs from that of $p\geq3$ in that correlation functions
can be obtained exactly from the properties of large random
matrices~\cite{Kost}.  

In this article, we will employ the functional-integral
formalism of~\cite{dom}.
To this end, we endow the
spin system with a relaxational dynamics,
\begin{equation}
\p_t\sigma_i=-\frac{\delta\beta\mathcal H}{\delta\sigma_i(t)}+\eta_i(t)
\label{equ:relax-dyn}
\end{equation}
where the $\eta_i$'s are independent white Gaussian noises with variance $2$. Following \cite{Kost,cugliandolodean}, it will prove useful to resort to the basis which diagonalizes the matrix of exchange couplings. The eigenvalues $\{J_\mu\}_{\mu=1,\ldots,N}$ of the $N\times N$ matrix $(J_{ij})_{i,j=1,\ldots,N}$ are distributed according to Wigner semi-circle law,
\begin{equation}\label{dis}
\rho(J_\mu)=\frac{1}{2\pi }\sqrt{4-J_\mu^2}
\end{equation}
Denoting by $\phi_\mu$ the spin coordinates in the basis that in which 
$(J_{ij})$ is diagonal, the Hamiltonian simplifies into
\begin{equation}\label{hamdiag}
\beta \mathcal{H}=-\frac{\beta}{2N}\sum_{\mu}J_{\mu}\phi_\mu^2 +\frac{u}{N}\left(\sum_{\mu}\phi_\mu^2\right)^2
\end{equation} 
and the equation of motion now reads
\begin{equation}\label{eqevolphimu}
\ppt\phi_\mu=
\beta J_\mu\phi_\mu+4u\phi_\mu\frac{1}{N}\sum_\nu\phi_\nu^2+\eta_\mu(t)
\end{equation}
where the $\eta_\mu$'s are independent white Gaussian noises with variance $2$. 

\subsection{Ferromagnetic model}

It been remarked that the $p=2$ spin glass resembles a ferromagnet 
`in disguise'~\cite{dedominicisgiardina}.
To illustrate this, we also consider a ferromagnetic model whose
Hamiltonian $\mathcal{H}_\mathrm{FM}$ is given by
\begin{equation}\label{hamfm}
\beta \mathcal{H}_\mathrm{FM}=
  -\beta\sum_{\langle ij\rangle}\sigma_i \sigma_j 
  +\frac{u}{N}\sum_{i,j}\sigma_i^2 \sigma_j^2 \;.
\end{equation} 
where the first sum runs over nearest neighbours on a $d$-dimensional
(hyper)-cubic lattice, but the $u$-term retains interactions
between all sites.  (Thus, the model contains
infinite-ranged couplings, as in the spherical ferromagnet.)
The analogues of the co-ordinates $\phi_\mu$ in this model are 
the Fourier transformed spin co-ordinates $\phi_{\bm{k}}$ where
$\bm{k}=(k_1,\dots,k_d)$ is the wave vector.  The 
eigenvalues of the matrix coupling the spins 
are $E_{\bm{k}}=\sum_{r=1}^d \cos k_r$.  The resulting equation
of motion is then
\begin{equation}\label{eqevolfm}
\ppt\phi_{\bm{k}}= \beta E_{\bm{k}} \phi_{\bm{k}}
  +4u\phi_{\bm{k}} \frac{1}{N}\sum_{\bm{k}}|\phi_{\bm{k}}|^2+
\eta_{\bm{k}}(t)
\end{equation}
where the $\eta_{\bm{k}}$ are independent Gaussian noises
as before.

In $d=3$ the distribution of the eigenvalues $E_{\bm{k}}$ 
is $\rho(E_k) \approx (2\pi^2)^{-1}\sqrt{d-E_k}$ when $|\bm{k}|$ is small.  Similarly, in the spin glass of~\eqref{ham},
the density of eigenvalues scales as $\rho(J_\mu) \approx
\pi^{-1}\sqrt{(2-J_\mu)}$ for $J_\mu$ close to 2.  
We will find that
the phase transitions in the models depend on the scaling
of the eigenvalue density near these points, and hence
that phase transitions in 
the $d=3$ ferromagnet and the $p=2$ spin glass are related
to each other, and have the same scaling exponents.

\subsection{Symmetry-breaking fields}
\label{sec:fields}

Below its transition temperature,
the ferromagnetic model spontaneously breaks the global
symmetry $\sigma_i \to -\sigma_i$.  To clarify the behaviour
in the ordered phase, it is convenient 
to introduce a magnetic
field: we take ${\cal H}_\mathrm{FM}
\to {\cal H}_\mathrm{FM} - h\sum_i\sigma_i$ in~\eqref{hamfm}.
The equation of motion becomes
\begin{equation}
\ppt\phi_{\bm{k}}= \beta ( E_{\bm{k}} \phi_{\bm{k}} 
 +  h\delta_{\bm{k},0}\sqrt N)
  +4u\phi_{\bm{k}} \frac{1}{N}\sum_{\bm{k}}|\phi_{\bm{k}}|^2+
\eta_{\bm{k}}(t)
\end{equation}
In the presence of this field, the magnetisation 
$\mathcal{M}(t)=N^{-1} \sum_i \sigma_i = N^{-1/2}\phi_{\bm{k}=0}$ 
acquires a finite expectation
value: in the ordered
phase, the magnetisation remains finite even in the limit of small
$h$, with a first-order phase transition at $h=0$.

In the spin glass model, the low temperature phase breaks the same
global symmetry, but the order parameter is not the magnetisation. 
Instead, the co-ordinate $\phi_\mu$
corresponding to the largest eigenvalue of $J_{ij}$ becomes
macroscopically occupied.  We assign the label 
$\mu=0$ to this eigenvector.  Then, by analogy with the ferromagnetic
case, we can introduce an analogous `staggered field' to the 
model of~\eqref{ham}.  The equation of motion becomes
\begin{equation}
\ppt\phi_\mu=\beta (J_\mu\phi_\mu
   + h_\mathrm{s} \delta_{\mu,0}\sqrt{N}) +
4u\phi_\mu\frac{1}{N}\sum_\nu\phi_\nu^2+\eta_\mu(t)
\end{equation}
The analogue of the magnetisation ${\mathcal M}(t)$ is the
staggered magnetisation ${\cal M}_\mathrm{s}(t)=N^{-1/2} \phi_{\mu=0}$.
In the
low temperature phase of the $p=2$ spin glass, the expectation
of $\mathcal{M}_\mathrm{s}(t)$
tends to a finite
value as $h_\mathrm{s}$ tends to zero, with a first-order
phase transition at $h_\mathrm{s}=0$.

\subsection{Thermodynamic formalism}

Ruelle's thermodynamic formalism involves a statistical mechanical analysis of the trajectories that a system follows through configuration space.
Let a {\it history} be a particular time-realization that the system has visited 
over a given time interval.  Consider an ensemble of histories constructed by 
fixing their dynamical activity $K(t)$.  Here, the dynamical activity 
is a history-dependent observable, extensive both in space and time, 
expressing the amount of activity within the history.
In a typical inactive history, the spins remain frozen 
in a locally ordered state; 
in an active history, spins fluctuate randomly between up and down states.
For both spin-glass and ferromagnetic models, a simple local observable 
consistent with this definition of activity 
is 
\begin{align}
K(t)=-\frac{1}{2}\sum_j\int_0^t\dd t\;\sigma_j^2(t) 
   =-\frac{1}{2}\sum_\mu\int_0^t\dd t\;\phi_\mu^2(t)
\label{act}
\end{align} 
Histories with $K$ close to 0 are the most active ones and are typically
associated with disordered states;
histories with large negative $K$ are inactive and are associated either
with local or global ordering.

While an ensemble of trajectories with fixed $K$ is natural from a physical
point of view, our theoretical methods require a change of ensemble, to one
in which the average activity is fixed.  (To draw an analogy with equilibrium 
statistical mechanics, we are transforming from a microcanonical to a canonical
ensemble.)
To fix the average activity, we apply a field $s$
that is conjugate to $K(t)$. 
While we are as yet unable to endow $s$ with an experimentally-realizable 
physical meaning, ensembles of histories with finite $s$ provide 
a valuable theoretical tool, which allow us to probe the 
histories that the system follows.   
The ensemble with $s=0$ is simply
the (unbiased) ensemble of trajectories for the system: ensembles
with $s>0$ are less active than the unbiased ensemble while those
with $s<0$ are more active.

In the following, we will evaluate the partition function  
\begin{equation}\label{defZdyn}
Z(s,t)=\langle \ee^{-s K}\rangle_0
\end{equation}
which is simply the generating function for the activity.  Here and throughout,
we use $\langle\cdot\rangle_0$ to denote an average of the
(unbiased) relaxational dynamics
over all possible time realizations, which means an average over the noises
$\eta_i$.  We also consider averages of a generic observable
$A$ in the biased ensemble parameterised by $s$, which we write as 
$\langle A \rangle_s \equiv \lim_{t\to\infty} Z^{-1}(s,t)
\langle A e^{-sK} \rangle_0$

We also define a dynamical free energy
\begin{equation}
\psi(s)=\lim_{t\to\infty}\frac{\ln Z(s,t)}{t}.
\end{equation}
It follows that 
\begin{equation}
\langle K \rangle_s = -Nt\frac{\mathrm{d}\psi}{\mathrm{d}s}
\end{equation}
With these definitions, singularities in $\psi(s)$ are the dynamical
phase transitions of the system.
Discontinuities in the derivatives of $\psi(s)$ will 
correspond to phase transitions between active and inactive
phases.
By analogy with equilibrium statistical mechanics, transitions
with a jump in $\langle K\rangle_s$ are termed `first-order' or
`discontinuous'; otherwise the transition is termed `continuous'.  
More specifically,
if $\langle K \rangle_s$ is continuous and the second
derivative is discontinuous 
then we refer to the transition as `second-order'.

\subsection{Functional integral formulation}
\label{sec:func-int}

To evaluate dynamical observables such as $\langle K\rangle_s$, 
we use the Janssen-De Dominicis functional-integral formulation~\cite{dom},
as in earlier studies such as Ref.~\cite{kirk}. 
The relaxational dynamics for the spin ${\phi}_\mu(t)$ is given by 
\eqref{eqevolphimu}. Using a functional integral representation the dynamical, 
$s$-dependent partition function introduced in \eqref{defZdyn} becomes
\begin{eqnarray} \label{part}
Z(s,t) =\int \D \phi \D \bar{\phi} \exp\left[ -\int_0^t\!\dd t' L(t') \right] 
\end{eqnarray}
where, omitting time-dependence for brevity,
\begin{equation}
L = 
\sum_\mu  \bar{\phi}_\mu\left(\partial_{t'} {\phi}_\mu -\beta
  J_\mu{\phi}_\mu
+\frac{4u}{N} \sum_\nu \phi_\nu^2{\phi}_\mu \right) 
-\bar{\phi}_\mu^2  - \frac{s}{2}{\phi}_\mu^2.
\end{equation}
(We consider the spin-glass model of~\eqref{ham} and we
have set the staggered field $h_\mathrm{s}=0$; other cases
will be discussed below.)

Due to the infinite-ranged interactions in the term proportional
to $u$, this model may be reduced to a quadratic form.  Details
are given in appendix~\ref{app:saddle}.  The result
is that the 
partition function becomes 
\begin{equation} \label{partfinal}
Z(s,t)=
\int \D \phi \D \bar{\phi} 
\exp\left[ 8u\bar{\chi}\chi Nt - \int_0^t \dd t' L_\mathrm{aux}(t') \right]
\end{equation}
with
\begin{equation}
L_\mathrm{aux} = \sum_\mu
\bar{\phi}_\mu(\partial_{t'} +4u\chi -\beta J_\mu)\phi_\mu 
-\bar{\phi}_\mu^2-\left(\frac{s}{2}-4 u \bar{\chi}\right)\phi_\mu^2
\end{equation}
where the parameters $\chi$ and $\bar\chi$ must be
determined self-consistently, through  
\begin{equation}
 \chi  = 
\int \dd J_\mu\, \rho(J_\mu)
\frac{1}{\sqrt{(4 u\chi-\beta J_\mu)^2-2(s-8 u\bar{\chi})}}
\label{sus}
\end{equation}
and
\begin{equation}
2\bar\chi + 1
=
\int\dd J_\mu\, \rho(J_\mu)\frac{4 u \chi-\beta J_\mu}{\sqrt{(4 u\chi-\beta J_\mu)^2-2(s-8 u\bar{\chi})}}
\label{susb}
\end{equation}

In addition, it follows from the definition of $\chi$ that
\begin{equation} \label{equ:chiK}
\chi = -2\langle K \rangle_s
\end{equation}
so that solving the self-consistency equations~\eqref{sus} and
~\eqref{susb} leads directly to the activities of the relevant
phases.  Finally, we note that the derivative
\begin{equation}
\frac{\mathrm{d}\chi}{\mathrm{d}s} = \langle ( K - \langle K \rangle_s )^2 
\rangle_s
\end{equation}
gives the fluctuations of the activity.

Physically, we have shown that the dynamical correlation functions
of the original model~\eqref{ham} are the same as those
of the auxiliary quadratic system~\eqref{partfinal}.  
Noting that $\rho(J_\mu)$ is
finite only for $-2<J_\mu<2$, the integrals in (\ref{sus})
and (\ref{susb}) are well-defined as long as
\begin{align}\label{cond}
(4 u\chi-2\beta)^2-2(s-8u\bar{\chi})>0
\end{align}
As long as this condition
is fulfilled, then the system is in a paramagnetic disordered phase.
On the other hand, if the denominator of~\eqref{sus} vanishes at $J_\mu=2$
then the mode associated with this eigenvalue may become macroscopically
populated.

\subsection{Symmetry-breaking fields}
\label{sec:fields_int}

As discussed in Sec.~\ref{sec:fields}, it is also useful
to consider the effects of symmetry-breaking fields $h_\mathrm{s}$
and $h$ on these systems.  Following
the analysis of the previous section, the symmetry-breaking fields 
lead to linear terms in $L_\mathrm{aux}$.
In general, the fluctuating magnetisation
${\cal M}_\mathrm{s}=N^{-1/2}\phi_{\mu=0}$
and its response field 
$\bar{\cal M}_\mathrm{s} = N^{-1/2}\bar\phi_{\mu=0}$
both have finite expectation values which we denote by
$m_\mathrm{s}$ and  $\bar{m}_\mathrm{s}$ respectively.
Evaluating these expectation values in the auxiliary model,
we arrive at
\begin{align}
\label{equ:ms}
m_\mathrm{s} &= 
   \beta h \frac{ 4u\chi - 2\beta }
                       { (4u\chi-2\beta)^2 - 2(s-8u\bar\chi) }
\\
\bar{m}_\mathrm{s} &=  
   \beta h \frac{ s-8u\bar\chi }
                       { (4u\chi-2\beta)^2 - 2(s-8u\bar\chi) }.
\end{align}
Self-consistency in the presence of the field
leads to modified saddle point
equations for $\chi$ and $\bar\chi$:
\begin{align}
\label{sus_h}
\chi&=m_\mathrm{s}^2 + 
\int \frac{ \dd J_\mu\, \rho(J_\mu)}
          {\sqrt{(4 u\chi-\beta J_\mu)^2-2(s-8 u\bar{\chi})}} 
\\
\label{susb_h}
 2\bar{\chi}+1&=2m_\mathrm{s} \bar{m}_\mathrm{s} + 
\int\frac{ \dd J_\mu\, \rho(J_\mu)(4 u \chi-\beta J_\mu)}
    {\sqrt{(4 u\chi-\beta J_\mu)^2-2(s-8 u\bar{\chi})}}.
\end{align}
 
Finally we note that while we have considered
the spin-glass model of~\eqref{ham}
throughout Secs.~\ref{sec:func-int}
and~\ref{sec:fields_int}, the equations for the
ferromagnetic model can be obtained by applying
the simple replacement 
$(\mu,J_\mu,m_\mathrm{s},\bar{m}_\mathrm{s})
 \to (\bm{k},E_{\bm{k}},m,\bar{m})$ throughout these
sections, where $\bar{m}=N^{-1/2}\bar\phi_{\bm{k}=0}$.

\section{Description of the phase diagram}
\label{sec:results}

\subsection{Overview}

\begin{figure}
  \includegraphics[width=7.5cm]{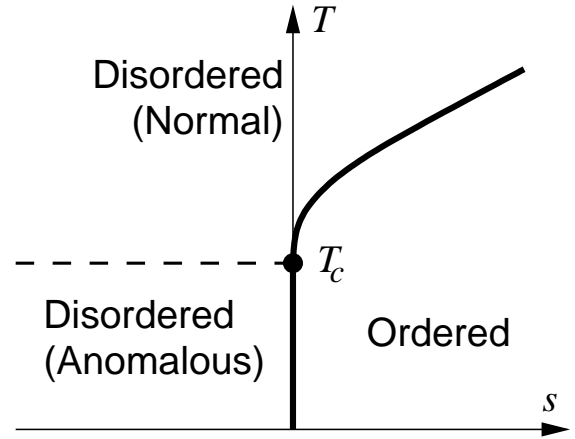}
  \caption{`Phase diagram' associated with the dynamic
free energy $\psi(s)$ of the spin glass model.  
The critical point of the model
is at $s=0$ and $T=T_c$.  The heavy solid line is a second
order phase boundary between ordered and disordered phases. 
The dashed line is a crossover within the disordered phase.
In the high temperature (normal) regime then the response
to a staggered field $\hs$ is positive; the low
temperature (anomalous) regime is characterised by a negative
response to this field.  }
\label{diagram}
\end{figure}

The dynamical phase diagram for the $p=2$ spin glass of
\eqref{ham} is shown
in Fig.~\ref{diagram}, for $h_\mathrm{s}=0$.  
The phase diagram of the ferromagnetic
model of~\eqref{hamfm} at $h=0$ has the same form.  
The axis $s=0$ corresponds
to the unbiased relaxational dynamics of~\eqref{equ:relax-dyn}.  
On this axis,
the spin-glass system has a second-order phase transition at 
\begin{equation} \label{Tc}
T_c=\frac{1}{\sqrt{2u}}.
\end{equation}
Below $T_c$, the mode with lowest eigenvalue is macroscopically populated,
and the system is ordered: the (staggered) magnetisation  
remains finite as the (staggered) field is reduced to zero.

We will show that the effect of positive $s$ is to promote 
ordering in the system, consistent with the expectation that ordered
phases are less active that disordered ones.  
As $s$ increased from zero, the second-order transition
between active and inactive phases moves to a higher temperature: we
have an increasing function $T_c(s)$, with $T_c(0)=T_c$ being
the thermodynamic transition temperature at $s=0$.

The effect of negative $s$ is to reduce ordering in the system, thus
increasing the activity: at temperatures
above $T_c$, the dynamical free energy $\psi(s)$ has 
no singularities for $s\leq0$
and $K$ increases smoothly as $s$ is decreased from zero.
We also find that no ordered phases
are possible for $s<0$: 
the condition~\eqref{cond} is always satisfied when and 
$\ms$ vanishes in the limit of small $\hs$.
Thus, for $0<T<T_c$,
behaviour of the model as $s\to0^+$ coincides with the 
ordered phase that is found at $s=0$, but
the behaviour is different for all $s<0$.  This signals the presence
of a phase boundary at $s=0$. 
The same effect is observed in the 
ferromagnetic model of~\eqref{hamfm}.  

We now show how this phase diagram is obtained
from the solutions to the self-consistency 
equations~\eqref{sus} and~\eqref{susb}.  We calculate
the saddle point average $\chi$ as a function of $s$
and the other parameters of the model: this gives the activity
of the phases of the model through~\eqref{equ:chiK}.  
In addition, we also calculate
the ($s$-dependent) staggered magnetisation $\ms$ which gives
additional insight into the phases of interest.  

\subsection{Unbiased dynamics ($s=0$)}

We begin with the unbiased ($s=0$) behavior of the model,
the derivation of which is identical to that of
Kosterlitz~{\it et al}~\cite{Kost} for the 
spherical version of this model.  
Equations (\ref{sus}) and (\ref{susb}) can be solved at $s=0$ 
with the result that $\bar{\chi}=0$, as required by causality, 
and that
\begin{equation}
\chi=\frac{1}{N}\sum_\mu\frac{1}{4u\chi-\beta J_\mu}
\end{equation}
If $\chi>\frac{\beta}{2u}$ then we can approximate the sum over the eigenvalues 
$J_\mu$ by an integral over the distribution $\rho(J_\mu)$, with the result 
\begin{equation}
\chi=\frac{1}{\beta^2}\left( 2u\chi-\sqrt{4 u^2\chi^2-\beta^2} \right)
\end{equation}
the solution of which is given by
\begin{equation}
\chi=\frac{1}{\sqrt{4u-\beta^2}}, \qquad
\beta<\beta_c,\;\;
\end{equation}
where $\beta_c=1/T_c=\sqrt{2u}$ consistent with~\eqref{Tc}.

However, for $\beta >\beta_c$, the 
mode with $J_\mu=2$ becomes macroscopically occupied, as described above. 
We therefore have 
\begin{equation}
\chi=\frac{\beta}{2u}, \qquad \beta>\beta_c.
\end{equation}
While the integral
in~\eqref{sus} is formally undefined, the ordered phase can be
studied either by introducing a finite staggered field $h_\mathrm{s}$
as discussed in Sec.~\ref{sec:fields_int}, 
or by solving~\eqref{sus} to $O(1/N)$.  In
either case, the staggered magnetisation $m_\mathrm{s}$ at zero field
is
\begin{equation}
\ms = m_0 \equiv \sqrt{\frac{T_c^2 - T^2}{T}}
\end{equation}

We also note a property of the ordered phase that is peculiar
to exactly soluble soft spins models such as the spherical
model~\cite{kac}.  The susceptibility associated
with the magnetisation, 
\begin{equation}
m^{(2)} = \langle ({\cal M}(t) - \ms)^2 \rangle_s  
\end{equation}
diverges as $\hs\to0$ for the unbiased dynamics ($s=0$) and
all $T<T_c$.  (This can be verified by evaluating
$\langle \phi_0^2 \rangle$ in the auxiliary
model.) This is in contrast to the usual situation
in critical phenomena where $m^{(2)}$ is finite for zero field and
$T<T_c$, diverging only at the critical point.

\subsection{Ordered phase, $s>0$}
\label{sec:ordered}

We now turn to the ordered phase for positive $s$.  
To simplify the analysis, we introduce
reduced variables:
\begin{eqnarray}
X(s) &=& 4 u T \chi(s) \\
Y(s) &=& 2T^2[s-8 u \bar{\chi}(s)] \;,
\end{eqnarray} 
where we explicitly indicate the $s$-dependence of $\chi$ and $\bar\chi$.
We see that $X(s)=\lambda^*$: physically, 
we identify the quantity $(-\lambda^* \phi_\mu)$
in~\eqref{lagrange} 
as the constraint force on mode $\mu$ that arises from the $u$-term 
in~\eqref{ham}
and suppresses configurations with extreme values of $\phi_\mu$.  We also
identify $Y(s)$ as a renormalised field $s$ for the auxiliary system.
It is easily verified that while the dynamical free energy $\psi$
depends on four parameters $(s,\beta,u,\hs)$, the properties
of the auxiliary model depends only on $(Y,\beta,X,\hs)$.
Our strategy will be to determine properties of the auxiliary model
in terms of $X$ and $Y$ and then to 
find the relations between $(X,Y)$ and the bare parameters of the model.

In the presence
of a staggered field $h_\mathrm{s}$, we have from~\eqref{sus_h} that
\begin{equation}
m_\mathrm{s} = \hs \frac{X-2}{(X-2)^2 - Y}
\end{equation}
For $Y>0$, spontaneous symmetry breaking occurs if the
denominator vanishes as $h_\mathrm{s}\to0$, as
\begin{equation}
\label{equ:XY-ord}
X=2 + \sqrt{Y} + O(\hs).
\end{equation}
Working at small $\hs$, we then take the zeroth order 
terms in~\eqref{sus_h}, arriving at
\begin{equation}
\label{equ:chiXY}
 \frac{2+\sqrt{Y}}{4uT} = \ms^2 + T \int\mathrm{d}J \rho(J) \frac{1}{
\left[ 2(2-J)\sqrt{Y}
 + (2-J)^2 \right]^{1/2}}
\end{equation}
This allows us to obtain 
$\ms^2 = m_0^2 + \frac{2\sqrt2 T}{\pi} Y^{1/4} + O(Y^{1/2})$.

Finally, we must relate the renormalised field $Y$ to the bare
field $s$.  Again working at zeroth order in $\hs$, \eqref{susb_h} becomes:
\begin{align}
 \frac{2s-Y\beta^2}{8u}& = \beta \ms^2 \sqrt{Y} - 1 + 
  \nonumber\\ 
& \int\mathrm{d}J \rho(J) 
\frac{2-J+\sqrt{Y}}{
\left[ 2(2-J)\sqrt{Y}
 + (2-J)^2 \right]^{1/2}}
  \label{equ:Ys-ord}
\end{align}
Taking $s$ small and positive, the solution has small positive $Y$.
More specifically,
the first term on the right hand side of~\eqref{equ:Ys-ord}
dominates as $Y\to0$, leading to
$\sqrt{Y}=\frac{s}{2(\beta^2-2u)}$.
[The same result can be obtained by working at $\hs=0$ and
considering carefully the limit of large-$N$.  The analogue
of~\eqref{equ:XY-ord} is $X=2+\sqrt{Y}+O(1/N)$ and we allow for a finite
values of $\ms$ and $\bar\ms$ when solving~\eqref{sus_h} and~\eqref{susb_h}.
The remainder of the analysis follows.]

Taking everything together, 
in the limit $\hs\to0^+$ and for $0< s\ll (T-T_\mathrm{c})$, the leading
behaviour is 
\begin{align}
\label{equ:chiFM}
\chi &\approx \frac{\beta}{2u} + \frac{s\beta}{8u(\beta^2-2u)}
\\
\ms^2 &\approx m_0^2 + \frac{2}{\pi\beta} \sqrt{\frac{s}{
\beta^2-2u}}
\end{align}
Physically, we can see that the $s=0$ axis in the phase
diagram of Fig.~\ref{diagram} is singular, but that both $\chi$
and $\mathrm{d}\chi/\mathrm{d}s$ are finite, so that fluctuations
of the activity
$K(t)$ remain finite as $h\to0$.  However, as for the
case $s=0$, the fluctuation
$m^{(2)}$ diverges as $h\to0$, for all cases where the spin-reversal
symmetry is spontaneously broken.

\subsection{High temperature regime}

We now turn to temperatures above the critical temperature, $\beta<\beta_c$.
We treat 
$s$ perturbatively in
(\ref{sus}) and
(\ref{susb}), arriving at
\begin{align}
\chi=\frac{1}{\sqrt{4 u-\beta^2}}+
  s\frac{1}{8(2 u-\beta^2)\sqrt{4 u-\beta^2}} 
\label{ci}
\end{align}
noting also that
\begin{align}
Y=2s\frac{2 u-\beta^2}{\beta^2(4 u-\beta^2)}\;.\label{cic}
\end{align}
We also notice that these solutions satisfy 
$(X-2)^2-Y>0$ at small $s$.

As the temperature is lowered towards $T_c$, we see
that $\mathrm{d}\chi/\mathrm{d}s$ diverges.  This again signals
the second-order transition to the ordered phase.
Indeed, it can be shown that this transition to the ordered phase
moves to a higher temperature for $s>0$.  
At the critical point,
we have $(X-2)^2=Y$ and $\ms=\bar{\ms}=0$.  With these
conditions, we can use~\eqref{equ:chiXY} and~\eqref{equ:Ys-ord} to 
derive the phase boundary for positive $s$.
In the limit
$\beta\to\beta_c^-$, the system is ordered for $s>s_c$, where
\begin{equation}
\label{equ:sc}
s_c(\beta)\simeq\frac{2\pi^2}{3\beta_c}(\beta_c-\beta)^3
\end{equation}
which holds to leading order in $\beta_c-\beta$.
This function gives the phase boundary in Fig.~\ref{diagram},
and its inverse gives the function $T_c(s)$ discussed above.

This completes our analysis of the phase diagram for $s\geq0$.  
There is an ordered phase separated from a paramagnet by a
second-order phase transition.  Loosely, the effect of positive
$s$ is simply to stabilise the ordered phase, so that spontaneous symmetry
breaking takes place at a higher temperature.

\subsection{Anomalous paramagnetic regimes}
\label{sec:anom}

We now take $s<0$ but we remain in the low temperature regime with 
$\beta>\beta_c$.  Working in terms of the reduced variables $X,Y$,
we take $Y<0$ so that we have
\begin{equation}
(X-2)^2-Y>0\;,
\end{equation}
and the integrand of~\eqref{susb} is finite for $-2<J_\mu<2$.
To make progress with the integrals in (\ref{sus},\ref{susb}): 
we define
\begin{eqnarray}
I_1(X,Y)=\int \dd J_\mu \rho(J_\mu)\frac{1}{\sqrt{(X-J_\mu)^2-Y}}
  \label{equ:i1} \;.\\
I_2(X,Y)=\int \dd J_\mu \rho(J_\mu)\frac{(X-J_\mu)}{\sqrt{(X-J_\mu)^2-Y}}\;
  \label{equ:i2}.
\end{eqnarray}
We will consider the limit $Y\to0^-$, in which the solution
to~\eqref{susb} is $X\to2^-$.  
The relevant limit is $0< (-Y)\ll
(2-X)^2\ll 1$.  
Writing $y'=-Y$ and $x'=2-X$, and
after some manipulations presented 
in appendix \ref{nontrivial}, we obtain:
\begin{align} 
& I_1=\pi^{-1}\sqrt{x'}\ln{(4x'^2/y')}+1+\frac{y'}{2 x^{\prime 3/2}}+O(x'^{1/2}) \label{I1}\\
&I_2=1-\frac{4(x')^{3/2}}{3\pi}+O(y' x'^{-1/2})+O(x'^{5/2})\label{I2}
\end{align}
The self-consistent equation~\eqref{sus} takes the form
\begin{equation}\label{x'}
\frac{\beta^2}{4 u}(2-x')=1+\frac{\sqrt{x'}}{\pi}\ln{\frac{4 {x'}^2}{y'}}
  +o(1)\;,
\end{equation}  
where we define $o(1)$ terms as quantities which vanish for
$y\ll x'^2 \ll 1$ .
Recalling that $X=2-x'=4uT\chi$, we have
to leading order
\begin{equation}
\chi\approx
\frac{\beta}{4 u}\left(2-\frac{\pi^2}{4 u^2}\frac{(\beta^2-2u)^2}{\ln^2{(-1/Y)}}\right) \;.
\end{equation}
Then, substituting for $x'$ in (\ref{I2}) and recalling~\eqref{susb}, 
we have 
\begin{equation}
\bar{\chi}\approx-\frac{\pi^2}{12u^3}\frac{(\beta^2-2 u)^3}{\ln^3{(-1/Y)}}
 \;.
\end{equation}
Noting that $y'=-2T^2(s-8 u \bar{\chi})$ we see that when
$s\rightarrow 0$, $\bar{\chi}\gg y'$, so that $s\simeq 8u
\bar{\chi}+o(1)$. This simplifies the expressions for $\chi$ and 
$\bar{\chi}$ which become:
\begin{align}
\label{equ:chi-anom}
\chi&\approx\frac{\beta}{2u}\left[1-\left(\frac{3\pi}{16 u}\right)^{\frac{2}{3}}(-s)^{\frac{2}{3}}\right]\\
\bar{\chi}&\approx\frac{s}{8u}
\end{align} 
which hold at leading order in $s<0$ and for $\beta>\beta_c$.

We refer to the phase with $s<0$ and $T<T_c$ as an
anomalous disordered phase.
To understand this terminology, it is useful to consider the
linear response to the field $\hs$, which is
given by~\eqref{equ:ms}.   Since we have $0\ll -Y \ll (2-X)^2$ 
with $X>2$, this reduces to
\begin{equation}
\ms = \frac{-\hs}{2-X}
\end{equation}
We can see that the response to the staggered field is
a staggered magnetisation in the opposite direction (a diamagnetic
response).  In addition, it follows from~\eqref{equ:chi-anom}
that $\mathrm{d}\ms/\mathrm{d}\hs \sim -|s|^{-2/3}$ for $s\to0^-$.
That is, the diamagnetic response diverges.  Clearly such a response
would be impossible in the unbiased ($s=0$) ensemble due to
thermodynamic convexity arguments, but when considering ensembles
with finite $s$ then such arguments do not apply.

Finally, we consider the nature of the
phase transition between ordered and anomalous disordered
phases.  Comparing \eqref{equ:chiFM} and \eqref{equ:chi-anom}, 
we note that $\chi$ is continuous at $s=0$,
and hence that the activity $\langle K\rangle_s$ is continuous also.
Thus, we identify a continuous phase transition at $s=0$, consistent
with Fig.~\ref{diagram}.  Often, at continuous phase transitions,
one may identify a path between the phases along which the free
energy is analytic and which remains always near
the critical point.  (For example, in a ferromagnet one can move between 
ordered and disordered phases by applying a small field $h$, decreasing 
the temperature and then removing the field.)  
However, in the transition considered here, the (staggered) magnetisation
$\ms$ is zero for $s<0$ but has a finite limit as $s\to0^+$.  This
seems to preclude such a route around the critical point.
Further, evaluating $(\mathrm{d}\chi/\mathrm{d}s)$ indicates
that the fluctuations of the activity diverge as $s\to0^-$ but
remain finite as $s\to0$.   This also indicates the absence
of a path between the phases that is continuous near the transition.
We turn to this issue in the next section, where we also discuss 
the possibility of diverging length scales near this transition.

\subsection{Effect of the staggered field $\hs$ on disordered phases}
\label{sec:anom_fields}

\begin{figure*}
\includegraphics[height=4.2cm]{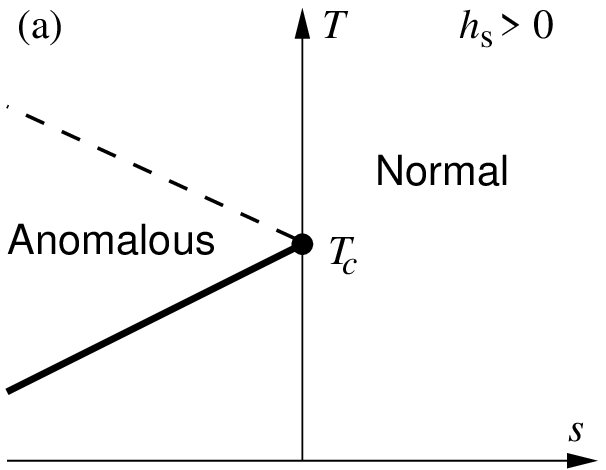}
\hspace{12pt}
\includegraphics[height=4.2cm]{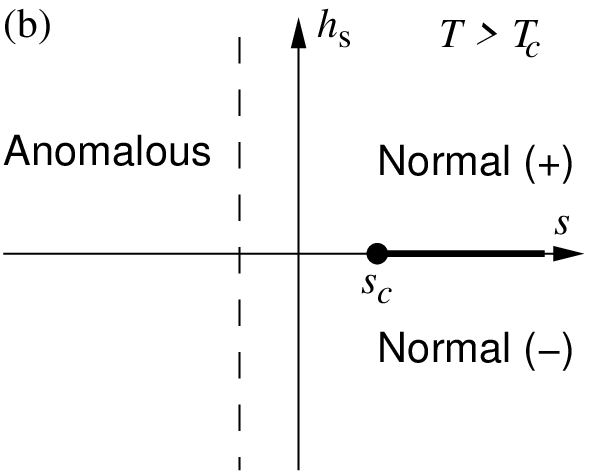}
\hspace{12pt}
\includegraphics[height=4.2cm]{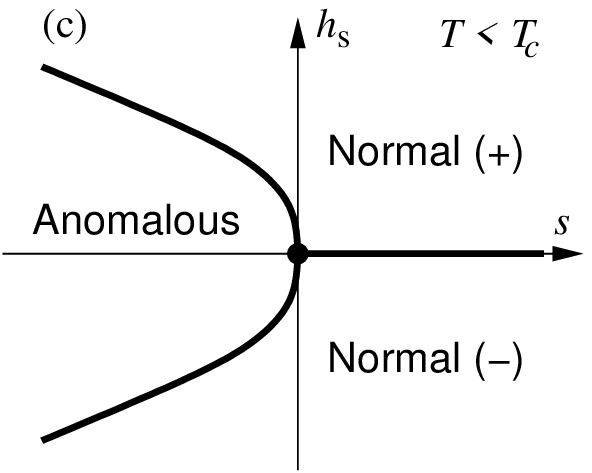}
\caption{(a) Proposed phase diagram for $\hs>0$.  The solid line
is a first-order phase transition at which the staggered
magnetisation and the activity $X$ are discontinuous.
It ends at a critical point at $T=T_c$ and $s=0$, but we
note that $\hs$ is finite at this critical point.   The dashed
line is a crossover at which the linear response to a staggered
field $\hs$ vanishes.  The dashed line is independent of $\hs$
while the solid line approaches the $s=0$ axis as $\hs\to0$.
(b)~Behaviour as a function of the field $\hs$ in the high
temperature regime.  For negative $s$, there is a crossover from normal
to anomalous response: at the crossover $X=2$ and
$Y$ satisfies \eqref{equ:ms0}.  For positive $s$, there is a critical
point at $s_\mathrm{c}$,
with spontaneous symmetry breaking for $s>s_\mathrm{c}$.
Close to $T_c$, the critical value of $s_\mathrm{c}$ is given
by \eqref{equ:sc}.
(c)~Behaviour as a function of 
$\hs$ in the low temperature regime. The solid line for $s<0$
is a first-order phase transition between states with
positive and negative response to the staggered field, while
the line for $s>0$ is the usual first-order
transition between spontaneously-ordered states.
}
\label{fig:phase_h}
\end{figure*}

To understand the behaviour near this phase transition 
in more detail, we introduce a finite staggered
field $h_\mathrm{s}>0$ and note that the crossover
between normal ($\ms>0$) and anomalous ($\ms<0$)
behaviour takes place at $\chi=\beta/2u$.
If we insist that $\chi$ take this
value, the self-consistency equation~\eqref{sus_h} becomes
\begin{equation}
\label{equ:ms0}
 \int \mathrm{d}J_\mu \rho(J_\mu)
 \frac{1}{\sqrt{(2-J)^2 - Y}}   = (\beta/\beta_c)^2
\end{equation}
For $\beta<\beta_c$ (high temperatures), this equation has a unique solution 
for $Y<0$, which signals a crossover from normal to anomalous
behaviour, at a value of $Y$ that is independent of the 
field $\hs$.  On the other hand, in the low temperature regime
$\beta>\beta_c$, (\ref{equ:ms0}) has no solutions. 

One may
verify that for large positive $s$, the solution of~\eqref{sus_h}
has $X\to0$, while 
for small positive $s$ we have from~\eqref{equ:chiFM} that
$X>2$.  Since \eqref{equ:ms0} establishes that, for low
temperatures, there
there are no values of $Y$ for which $X=2$, it follows
that the $s$-dependent value of $X$ has a jump
from a value greater than 2
to a value smaller than 2.  This is a discontinuous phase
transition from normal to anomalous states.  (The field 
$\hs$ is finite, so the concept of a spontaneously ordered
state is not useful.  However, the sign of the staggered
magnetisation is positive in the normal state and 
negative in the anomalous one.)  These arguments lead
us to propose the qualitative phase diagram shown 
in Fig.~\ref{fig:phase_h}.  The presence of
first-order transitions at finite $\hs$ explains
the unusual features of the order-disorder
transition at $s=\hs=0$: there are indeed no continuous
paths between the ordered and disordered states due
to the first order transition at finite $\hs$.

The nature of the scaling behaviour near $T_c$
in this model is clearly complicated,
depending qualitatively on the order in which
$s$, $\hs$ and $T-T_c$ are taken to zero, and
also on the signs of $s$ and $T-T_c$.  A
detailed investigation of these finite-$\hs$
transitions is beyond the scope of this paper.  
However, we can conclude that the order-disorder
transition at $\hs=s=0$ is second-order in that 
$\chi(s)$ is continuous, but that the spontaneous
staggered magnetisation $\ms$ goes discontinously
to zero at this transition.   We again emphasise
that all of this phenomenology is also present
in the ferromagnetic model of~\eqref{hamfm},
at least for $d=3$.  We expect the qualitative
features to also be present in higher dimensions.

In the ferromagnetic model, we can also consider
the correlation lengths of the various phases.
These appear through the $\bm{k}$-dependence
of the fluctuations:
\begin{equation}
S(\bm{k},s) \equiv \langle |\phi_{\bm{k}}|^2 \rangle_s 
 = \frac{T}{\sqrt{(X-E_{\bm{k}})^2 - Y}}
\end{equation}
Recall that $E_{\bm{k}}\leq d$ with equality if $\bm{k}=0$.

Two cases are of interest.  Firstly, if the denominator
vanishes at $\bm{k}=0$ as in the ordered states, the 
fluctuations of the spontaneous magnetisation diverge,
as described above.  Secondly, if $X<d$ as in the anomalous
phase then
$S(\bm{k})$ has a peak at a finite wave-vector $\bm{k}^*$
for which $E_{\bm{k}^*}=X$.
We interpret $1/|\bm{k}^*|$ as a characteristic
length scale for structures within this phase.  It is interesting
to note that this length scale diverges as $s\to0^-$ in the
anomalous phase, and that this is
 accompanied by a divergence in the fluctuations
of the spontaneous magnetisation [it may be easily
verified that $S(\bm{0},s\to0^-)$ is divergent since $X\to d^-$
and $Y\to0^-$ in this limit, by analogy with the $p=2$
spin glass].

\section{Interpretation}
\label{sec:interp}

We have considered in some detail the large deviations of the
dynamical activity in two soft-spin models.  We end with a
comparison with previous studies and with some comments
on the relation between the large deviations that we studied
and the phase behaviour of the models.

Based on the close relationship between the ferromagnetic
and spin glass models, the form of the phase diagram in
Fig.~\ref{diagram}
is perhaps not too surprising: a similar result was found 
in Ref.~\cite{thermo}  for the infinite-ranged (mean-field) Ising model.  
The equilibrium critical point leads to 
a dynamical phase transition at $s=0$ below the critical temperature. 
However, instead of being first order as in the fully-connected model, 
the phase transition we have found is second order.  Unlike the
fully-connected models, the ferromagnetic model of \eqref{hamfm}
has several diverging length scales, although the presence of 
a diverging length scale throughout the ordered phase may be 
a peculiarity of our particular model.  In any case, the diverging
length scale as $s\to0^-$
 within the anomalous phase seems to be a new feature
that merits further investigation.

In particular, the existence of the anomalous phase seems to be
linked to the existence of aging/coarsening solutions to the
relaxational dynamics of these models.  These solutions are characterised
by $\ms=0$ as in the anomalous phase and exhibit a length scale
that grows with the time that has elapsed since a quench from above
$T_c$.  It can be readily shown that if multiple solutions to the equations
of motion exist with different activity, then the field $s$ acts
to select the solutions with the larger ($s>0)$ or smaller ($s<0$) 
activity, leading to a transition at $s=0$.  
However, while the aging dynamics of the $p=2$ spin-glass
of \eqref{ham} can be solved~\cite{coarse} by a similar method to that 
of Cugliandolo and Dean~\cite{cugliandolodean}, we have not yet
established any clear connection between these dynamics and the 
anomalous disordered phases discussed here.  This too remains an area
for future study.

We also compare the results shown here with those obtained for
kinetically constrained models~\cite{lettres}.
In both cases, active and inactive phases coexist at the
$s=0$ axis.  However, there are two important differences. 
Firstly, in the kinetically constrained
models considered in~\cite{lettres}, the transition is first-order,
signalling the coexistence of active and inactive solutions to
the equations of motion.  These have been interpreted as 
``ergodic fluid'' and ``non-ergodic glass'' 
states~\cite{garr,merolle,lettres,ka-science}.
On the other hand, the continuous transition in the soft
spin models is second order: the anomalous phase is characterised
by a diverging correlation length that we have tentatively
attributed to the growing length scale associated 
with the aging behaviour of the system.  Taking the large-time
limit of the aging solution, the activity of the system approaches
that of the ordered state: the active (aging) and inactive
(ordered) phase are not separated by a gap in the activity,
unlike the kinetically constrained models.

Secondly, we emphasise that in the kinetically constrained models,
the $s=0$ axis of the phase diagram belongs to the active phase. 
Introducing any $s>0$ leads immediately to an inactive phase that
is qualitatively different from the unbiased state at $s=0$.  On
the other hand, in the models considered here, the $s=0$
axis belongs to the inactive (ordered phase): it is the introduction
of any $s<0$ that leads to an active phase that differs
from the unbiased steady state. 

Finally we note that 
contrary to its $p\geq 3$ counterparts, the thermodynamic
properties of the $p=2$ spin glass have only a single
transition temperature and do not display any kind 
of replica symmetry breaking.  Here, we have analysed
this problem by diagonalising the quadratic dynamical action:
a method that applies only for $p=2$.  However, the same
results can be verified using the replica trick  
and integrating out the disorder.  The application of such
methods to models with $p\geq3$ would provide further insight
into the behaviour of large deviations of the activity in 
`glassy' models.

\begin{acknowledgments}
We thank Juan Garrahan for many helpful discussions.
We also thank David
Chandler for his advice and encouragment in the early
stages of this work, during which time
RLJ was supported by the Office of Naval
Research through grant No.~N00014-07-1-068. 
We are grateful for financial support from the
Franco-British Alliance programme,
managed by the British Council and the French Foreign Affairs Ministry
(MAE).
\end{acknowledgments}

\begin{appendix}

\section{Saddle-point integration of the dynamical action}
\label{app:saddle}

Here, we show how the functional integral (\ref{part}) can be
cast in the form (\ref{partfinal}).  We define
\begin{equation}\label{chi}
{\cal X}(t)=\frac{1}{N}\sum_\mu {\phi}_\mu^2(t) \;,
\end{equation}
which is proportional to the growth rate of the activity $K$ defined in
\eqref{act}, and the related quantity:
\begin{equation}\label{chibar}
\bar{\cal X}(t)=\frac{1}{N}\sum_\mu\bar{\phi}_\mu(t) {\phi}_\mu(t)\;.
\end{equation}

Both $\cal X$ and $\bar{\cal X}$ 
are self-averaging quantities in the thermodynamic
limit of large $N$.   It is therefore convenient to constrain these
quantities with Lagrange multipliers and then to integrate over these
constrained quantities by a saddle-point method.
Writing the (time-dependent) Lagrange multipliers as
$\lambda, \bar{\lambda}$, Eq. (\ref{part}) becomes:
\begin{align}\label{lagrange}
&Z(s,t)=\int \D{\cal X} \D
\bar{\cal X}\frac{N\D\bar{\lambda}}{2\pi i}\frac{N\D\lambda}{2 \pi i}\mathcal{D}\phi\mathcal{D}\bar{\phi}\, \left[ \ee^{ -NS_0(t) - S_1(t) } \right]
\end{align}
with 
\begin{align}
S_0(t) &= \int_0^t\dd t' \left[ \bar{\lambda}(t'){\cal X}(t')
 +{\lambda}(t')\bar{\cal X}(t')+4 u
  {\cal X}(t')\bar{\cal X}(t') \right] 
\end{align}
and
\begin{align}
S_1(t) = \int_0^t\dd t' \sum_\mu \bigg\{ &\bar{\phi}_\mu 
  \left[\frac{\partial}{\partial t'} + \lambda(t') 
  -\beta J_\mu  \right]\phi_\mu  \nonumber \\
& - \left[\frac{s}{2} + \bar{\lambda}(t')\right] \phi_\mu^2 
  - {\bar\phi}^2_\mu
\bigg\}
\end{align}
where we again omit the dependence of the fields $\phi$ and $\bar{\phi}$ 
on the time $t'$, for brevity.

In the $N\rightarrow\infty$ limit, the integrals over $\cal X$,
$\bar{\cal X}$, $\lambda$ and $\bar{\lambda}$ can be carried out 
through a saddle point
approximation.  We replace ${\cal X}(t)$, $\bar{\cal X}(t)$, 
$\lambda(t)$ and $\bar\lambda(t)$ 
by their saddle-point values $\chi$, $\bar\chi$, $\lambda^*$
and $\lambda^*$.   In particular, differentiating the action $(N S_0 + S_1)$ 
with
respect to $\cal X$ and $\bar{\cal X}$, we arrive at
\begin{align}
\bar{\lambda}^* &= -4 u \bar{\chi}\\
{\lambda}^* &=-4u \chi\;.
\end{align}
Thus, performing the saddle-point integrals in~\eqref{lagrange}
leads to~\eqref{partfinal} in the main text.
Similarly, differentiating with respect to $\lambda$ and $\bar\lambda$
leads to the self-consistency equations (\ref{sus}) and (\ref{susb}).

\section{Computation of $\chi$ in the low-temperature disordered
  phase}\label{nontrivial}

Here we discuss the solutions of the self-consistency equations~\eqref{sus}
and~\eqref{susb} for $s<0$ and $T<T_c$.  
Using the notation of Sec.~\ref{sec:anom}, we can write the integrals
of Equs.~\eqref{equ:i1} and~\eqref{equ:i2} as
\begin{align}
I_1&=\int_{-4+x'}^{x'}\frac{\dd z}{2\pi}
  \frac{\sqrt{4(x'-z)-(x'-z)^2}}{\sqrt{z^2+y'}}\label{I1in}\\
I_2&=\int_{-4+x'}^{x'}\frac{\dd z}{2\pi}\frac{-z\sqrt{4(x'-z)-(x'-z)^2}}{\sqrt{z^2+y'}}\label{I2in}
\end{align}
Since we are at $\beta>\beta_c$ and $s<0$, we have $x',y'>0$. The small $s$
limit becomes the limit $y'\rightarrow 0$ but if we take the limit for $y'\rightarrow 0$ keeping $x'>0$ we
have $I_1(x',y'\rightarrow 0)\rightarrow \infty$, and the self-consistency
condition cannot be satisfied.  We therefore take both $x'$ and
$y'$ to zero together: we assume that $y'\ll x'^2$ which
can be verified \emph{a posteriori} through the solution 
to~\eqref{x'}.  The result is $x'\sim [\ln(1/y')]^{-2}$,
consistent with our assumptions.

We start by splitting the integral in
(\ref{I1in}) into three parts:
\begin{align}
I_1&= \int_{-x'}^{x'}\frac{dz}{2\pi}
  \frac{\sqrt{x'(4-x')}}{\sqrt{y'+z^2}}\nonumber \\
&+\int_{-x'}^{x'}\frac{dz}{2\pi}
  \frac{\sqrt{4(x'-z)-(x'-z)^2}-\sqrt{x'(4-x')}}{\sqrt{z^2+y'}}\; \nonumber \\
&+\int_{x'-4}^{-x'}
\frac{dz}{2\pi}\frac{\sqrt{4(x'-z)-(x'-z)^2}}{\sqrt{z^2+y'}}\;.
\end{align}
In the limit of $y'\ll x'^2$, the first
integral has a divergent contribution
$(\sqrt{x'}/\pi)\ln{(4{x'}^2/y')}$.
In the second and third parts we can 
take the limit $y'\rightarrow 0$ directly in the integrand.  If we
then take the limit of small $x'$ then we find that the
second integral vanishes as $O(\sqrt{x'})$
while the third approaches unity.  Thus we arrive
at~\eqref{I1}.

We now evaluate the $(y'/{x'}^2)\rightarrow 0$ limit of the expression
(\ref{I2in}). We use a similar method, spliting the integral into two terms:
\begin{align}\label{split}
I_2&=1+\nonumber\\
&\int_{-x'}^{x'}\frac{dz}{2\pi}\sqrt{4(x'-z)-(x'-z)^2}\left(
  \frac{-z}{\sqrt{z^2+y'}}-1\right)\nonumber\\
&+\int_{x'-4}^{-x'}\frac{dz}{2\pi}\sqrt{4(x'-z)-(x'-z)^2}\left(
  \frac{-z}{\sqrt{z^2+y'}}-1\right)\;.
\end{align}
In the first integral we introduce $w=z/x'$ and $\sigma=y'/{x'}^2$, thus 
arriving at
\begin{equation}
(x')^{3/2}\int_{-1}^1 \frac{dw}{2\pi}(
  \frac{-w}{\sigma+w^2}-1)\sqrt{4(1-w)-x'(1-w)^2)}\;.
\end{equation}
The leading behaviour of this quantity can be evaluated
by setting directly $\sigma=x'=0$ into the integral, so that
this contribution to $I_2$ is
$-4 (x')^{3/2}/(3\pi)[1+o(1)]$.

For the second integral in Eq. (\ref{split}), we 
have $z^2\geq x'^2$ so we expand the integrand in powers of $\sigma=y'/z^2$.
The leading
term vanishes and the second term is $O(y'/\sqrt{x'})$. Writing 
$(y'/\sqrt{x})=(x')^{3/2} (y'/x'^2)$, this term is smaller first
term in~\eqref{split} which is
$O((x')^{3/2})$. Thus we arrive at
\begin{equation}
I_2\approx 1-\frac{4(x')^{3/2}}{3\pi}\;.
\end{equation}
as given in Equ.~\eqref{I2} of the main text.

\end{appendix}


\end{document}